\documentclass[pre,onecolumn,notitlepage]{revtex4-1}
\usepackage{graphicx}
\usepackage{amssymb,amsthm,amsmath,physics,xcolor}

\newcommand{\beq}[0]{\begin{equation}}
\newcommand{\eeq}[0]{\end{equation}}

\begin{document}

\title{Interactions of solitary waves in the Adlam-Allen model}
\author{Boris A. Malomed$^{1,2}$}
\author{Panayotis G. Kevrekidis$^3$}
\author{Vassilis Koukouloyannis$^4$}
\author{Nathaniel Whitaker$^3$}
\author{Dimitrios J. Frantzeskakis$^5$}

\begin{abstract}
We study the interactions of two or more solitary waves in the Adlam-Allen
model describing the evolution of a (cold) plasma of positive and negative
charges, in the presence of electric and transverse magnetic fields. In
order to show that the interactions feature an exponentially repulsive
nature, we elaborate two distinct approaches: (a) using energetic
considerations and the Hamiltonian structure of the model; (b) using the
so-called Manton's method. 
We compare these findings with results of direct simulations, and identify
adjustments necessary to achieve a quantitative match between them.
Additional connections are made, such as with solitons of the Korteweg--de
Vries equation. New challenges are identified in connection to this model
and its solitary waves.
\end{abstract}

\maketitle

\affiliation{$^{1}$Department of Interdisciplinary Studies, Faculty of
Engineering, Tel Aviv University, Tel Aviv 69978, Israel\\
$^{2}$Instituto de Alta Investigaci\'{o}n, Universidad de Tarapac\'{a}, Casilla 7D, Arica, Chile}

\affiliation{$^{3}$Department of Mathematics and Statistics, University of
Massachusetts, Amherst MA 01003-4515, USA}

\affiliation{$^{4}$Department of Mathematics, Laboratory of Applied Mathematics
and Mathematical Modelling, University of the Aegean, Karlovasi, 83200
Samos, Greece}


\affiliation{$^5$Department of Physics, National and Kapodistrian University of
Athens, Panepistimiopolis, Zografos, Athens 15784, Greece}

\section{Introduction}

The field of nonlinear plasma physics has been a rich source of intriguing
problems for the dynamics of solitary waves in integrable and
nearly-integrable systems \cite{infeld,konosk}. In particular, the famous
work of Zabusky and Kruskal \cite{kz}, which initiated the explosion of
interest in solitons by showing that the continuum limit of the
Fermi-Pasta-Ulam-Tsingou model~\cite{fpu,fput} is the Korteweg-de Vries
(KdV) equation~\cite{kdv}, was a motivating theme for the work of Washimi
and Taniuti~\cite{tan}. The latter one demonstrated that small-amplitude
ion-acoustic waves in plasmas are also governed by the KdV model, hence
solitonic excitations may be expected in this setting too. However, as
indicated in the historical review of early studies of solitons \cite{allen3}%
, it was overlooked in the seminal works \cite{kz} and \cite{tan}, and in
the extensive literature initiated by them (see, e.g.,~\cite%
{infeld,Rem,mjarecent}), that solitary waves were discovered in plasmas well
before Refs. \cite{kz} and \cite{tan}. Indeed, a fundamental model put forth
by Adlam and Allen in 1958 and 1960~\cite{aa1,aa2} constitutes, arguably,
one of the earliest encounters of the realm of plasma physics with the
concept of solitary waves.

The analysis presented in Refs.~\cite{aa1} and \cite{aa2} concerns the
spatiotemporal evolution of the distribution of electrons and ions in a
magnetized plasma. In this setting, the spatial variation occurs only along
the $x$-direction, the electric field acts along the $\left( x,y\right) $
plane, being subject to the Faraday's and Gauss' laws, while the
(transverse) magnetic field acting along the $z$-direction obeys the Amp{%
\'{e}}re's law. The Newtonian spatiotemporal dynamics of a plasma consisting
of positive and negative charges is affected by the forces created by the
electric and magnetic fields. In this framework, starting from first
principles and utilizing well-established approximations, such as the
quasi-neutrality~concept, a reduced system for plasma dynamics under the
action of the electromagnetic field was derived \cite{aa1,aa2}. Ultimately,
the resulting Adlam-Allen (AA) system of partial differential equations
(PDEs) involved only two evolution equations for the (rescaled) magnetic
field $B$ and inverse density $R$, in the $(1+1)$-dimensional setting \cite%
{aa1,aa2} (see Ref. \cite{allen4} for a recent recount of the topic).

The AA system is the starting point of the present work. In particular, in a
recent study~\cite{PRE} this nonlinear model of plasma physics was
revisited, and key properties of its solutions, including solitary and
periodic waves, were examined. In addition to that, a connection of the AA
model to the KdV equation was established (see also Ref.~\cite{nairn})
through a multiscale reduction, and collisions of solitary waves were
briefly addressed. In the present work, we aim to study interactions of
solitary wave in the AA system in detail. It is well known that solitons in
the KdV equation repel each other, and exact multi-soliton solutions can be
obtained by means of the inverse-scattering transform (IST) method~\cite%
{mjarecent,whitham,leveque}. Studies of interactions of solitary wave in
non-integrable models, relying upon the identification of their pairwise
potential~\cite{interaction} or force~\cite{manton}, have been the subject
of numerous studies; see, e.g., Ref.~\cite{kivshar} for an early review of
relevant results.

On the basis of the reduction of the AA model to the KdV equation for weakly
supersonic speeds of the solitary waves~\cite{PRE}, it is natural to expect
repulsion between them in the AA system as well. However, the AA model does
not have the integrable structure of the KdV, which provides exact
multi-soliton solutions, therefore one needs to resort to asymptotic
techniques, such as ones based on the Lagrangian/Hamiltonian structure of
the model~\cite{interaction}, or others which directly address the system of
PDEs and related conservation laws~\cite{manton}. Here, we leverage both of
these approaches and conclude that they lead to the \textit{same}
conclusions for the repulsion of the solitary waves. We then go on to
corroborate analytical predictions by means of direct simulations.

The subsequent presentation is organized as follows. In Section~II, we
present the physical and mathematical basis of the above-mentioned setup,
including its Lagrangian and Hamiltonian structure and solitary-wave
solutions. In Section~III, we use the asymptotic form of the solitary waves
and the Hamiltonian of the AA system to address the tail-tail interaction
between solitary waves. In Section~IV, we compare the predictions to
numerical simulations and identify adjustments needed for a quantitative
match between them. In Section~V, we summarize our findings and highlight
some directions for future studies. Finally, in the Appendix we provide an
alternative systematic proof of our results for the interaction between
solitary waves, by means of the so-called \textit{Manton method}~\cite%
{manton,panos_inter}. A second Appendix offers a perspective on a different,
but also important type of interaction, namely that of a solitary wave with
a localized defect.


\section{The model, its properties, and solitary waves}

\subsection{Introducing the AA model}

The AA model introduced in Refs.~\cite{aa1,aa2} describes the wave
propagation in a cold magnetized collisionless electron-ion plasma. In
particular, the thermal motion is negligible in comparison to velocities of
the particles due to the wave motion, and collisions are also neglected due
to the fact that the collision frequencies are small (i.e., the mean time
between collisions is much larger than the time which an ion/electron spends
in the wave). Electrons and ions in the plasma are subject to the action of
the magnetic field applied in the $z$-direction, and there is an induced
electric field in the $y$-direction, while the assumption of the
quasi-neutrality is consonant with the presence of a weak electric field in
the $x$-direction. The latter is true as long as the electron plasma
frequency is much greater than the electron cyclotron frequency. Note that
such a setting may find applications both in fusion research and in studies
of astrophysical phenomena, such as the solar wind~\cite{solar}.

Adlam and Allen described how a large-amplitude stationary compressional
magnetic-field pulse (or a train of pulses \cite{PRE}) can exist and be
sustained in the collisionless plasma. In particular, adopting the
Lagrangian coordinate system (moving with the pulse), they have found a
nonlinear solution of such a model involving ions, electrons and the
electric and magnetic fields. This solution corresponds to accumulation of
the magnetic flux, which is sustained by the flow of the plasma across it.
The particles' velocities must be large enough, so that the ion Larmor
radius is larger than the effective width $D$ of the magnetic pulse, and the
electric field of the ions is able to pull the electrons across $D$. Then, $%
D $ turns out to be $\sim c/\omega _{p}$, i.e., on the order of the
colisionless skin depth (with $c$ being the speed of light and $\omega _{p}$
the plasma frequency), and the strength of the magnetic pulse depends on the
Alfv{\'{e}}n Mach number $M_{A}$, which lies in the interval of $1\leq
M_{A}\leq 2$ (see e.g. \cite{swanson,stix}).

The AA system can be expressed in the following dimensionless form \cite{PRE}
(see also Ref.~\cite{nairn}):
\begin{eqnarray}
R_{tt}+\frac{1}{2}\left( B^{2}\right) _{xx} &=&0,  \label{R} \\
B_{xx}-RB+R_{0}B_{0} &=&0,  \label{B}
\end{eqnarray}%
where the real functions $R(x,t)>0$ and $B(x,t)$ represent, respectively,
the inverse plasma density and the magnetic field, while constants $R_{0}$
and $B_{0}$ are the density and magnetic field strength in the undisturbed
plasma, respectively (note that the plasma is assumed to be initially
uniform and steady, i.e., $B=B_{0}$, and $R=R_{0}$, at $t=0$). These
constants also set the boundary conditions (b.c.) at infinity, i.e., $%
R\rightarrow R_{0}$ and $B\rightarrow B_{0}$ as $|x|\rightarrow \infty $;
notice that $R_{0}$ and $B_{0}$ are related by the following equation,
\begin{equation}
R_{0}\equiv B_{0}^{2}/C^{2},  \label{R0}
\end{equation}%
where $C$ is the characteristic speed of small-amplitude waves propagating
on top of the background solution $R=R_{0}$, $B=B_{0}$; details of the
derivation and scaling of the AA system can be found in Refs.~\cite%
{aa1,aa2,PRE,nairn}.

As mentioned above, Adlam and Allen have found a class of large-amplitude
hydromagnetic solitary waves which propagate in this setting. Their
analytical treatment was inherently nonlinear, due to the consideration of
finite-amplitude waves and self-localization effects. Therefore, this
treatment differs from that of linear waves commonly appearing in textbooks
on plasma waves (see, e.g., Refs.~\cite{swanson,stix}), according to which
the (linear) small-amplitude waves are considered as weak perturbations
propagating on top of a background equilibrium.

In this work, we also focus on the fully nonlinear version of the AA model.
However, the finite-amplitude solitary waves that we consider here share
their qualitative properties with small-amplitude fast magnetoacoustic Alfv{%
\'{e}}n modes, as obtained from the linear theory, under the approxomation
of the cold collisionless plasma~\cite{swanson,stix}. In fact, these are
compressional electromagnetic waves, propagating perpendicularly to the
background magnetic field. The particle motion in the waves is in the
direction transverse to the background magnetic field, with the electric and
magnetic fields of the wave oriented perpendicular and parallel to the
background magnetic field, respectively. The AA model describes the
self-localization of the waves along the $x$-direction perpendicular to the
background magnetic field (which lies along the $z$-direction). Furthermore,
the Faraday's law, in conjunction with the presence of the magnetic field,
accounts for the $y$-component of the electric field, while the $x$%
-component of the latter obeys the Gauss' law.

\subsection{Solitary waves}

It is convenient to eliminate the constant background from Eqs.~(\ref{R})-(%
\ref{B}, upon introducing the following definitions:
\begin{equation}
R\left( x,t\right) \equiv R_{0}+u(x,t),~B(x,t)\equiv B_{0}+w\left(
x,t\right) ,  \label{background}
\end{equation}%
where the fields $u$ and $w$ satisfy vanishing b.c. at infinity, namely $%
u,w\rightarrow 0$ as $|x|\rightarrow \infty $. Then, the respectively
transformed Eqs.~(\ref{R}) and (\ref{B}) read \cite{PRE}:
\begin{eqnarray}
u_{tt}+\left( \frac{1}{2}w^{2}+B_{0}w\right) _{xx} &=&0,  \label{u} \\
w_{xx}-R_{0}w-B_{0}u-uw &=&0.  \label{w}
\end{eqnarray}


As shown in Ref.~\cite{PRE} (see also Ref.~\cite{davis} for an earlier
similar analysis), Eqs.~(\ref{u}) and (\ref{w}) possess an exact
solitary-wave solution of the form:
\begin{eqnarray}
w_{\mathrm{sol}} &=&\frac{2B_{0}}{C}\frac{v^{2}-C^{2}}{C+v\cosh \left( \frac{%
B_{0}\sqrt{v^{2}-C^{2}}}{Cv}\xi \right) },  \label{wsol} \\
u_{\mathrm{sol}} &=&-\frac{1}{v^{2}}\left( B_{0}w_{\mathrm{sol}}+\frac{1}{2}%
w_{\mathrm{sol}}^{2}\right) ,  \label{usol} \\
\xi &\equiv &x-vt.  \label{xi}
\end{eqnarray}%
Here, $v$ is the solitary-wave's velocity, which takes values in the
interval of
\begin{equation}
C<v<2C.  \label{C}
\end{equation}%
The lower bound $C$ of $v$ is set by the necessary condition for the
existence of the homoclinic orbit that corresponds to the exact
solitary-wave solution (this homoclinic orbit occurs in the phase plane of
the dynamical system stemming from Eqs.~(\ref{u})-(\ref{w}) once
traveling-waves solutions are sought). In terms of the underlying physics,
this condition means that the nonlinear solitary waves propagate at speeds
higher than that of the linear-wave propagation in the system~\cite{PRE}. On
the other hand, the upper bound $2C$ for $v$ in Eq. (\ref{C}) follows from
the requirement that the (inverse) density $R$ must be positive definite.
While formal solutions exist past this threshold, they have no physical
meaning.

\subsection{The Lagrangian and Hamiltonian structure}

Here, we aim to reveal the Lagrangian/Hamiltonian structure of the AA
system. For this purpose, following Ref.~\cite{Chalmers}, it is relevant to
define potential $U(x,t)$ of field $u\left( x,t\right) $,
\begin{equation}
u\equiv \partial U/\partial x.  \label{potential}
\end{equation}%
The substitution of definition (\ref{potential})\ in Eqs.~(\ref{u}) and (\ref%
{w}) and subsequent integration of the former equation with respect to $x$
replaces Eqs.~(\ref{u}) and (\ref{w}) by the following equations:
\begin{gather}
U_{tt}+\left( \frac{1}{2}w^{2}+B_{0}w\right) _{x}=0,  \label{U} \\
w_{xx}-R_{0}w-B_{0}U_{x}-wU_{x}=0,  \label{wU}
\end{gather}%
where we have set the constant of integration (which, in principle, may be a
function of time) equal to zero, as per the assumption that $U(x)$ and $w(x)$
vanish as $|x|\rightarrow \infty $.

Next, it is straightforward to see that Eqs.~(\ref{U}) and (\ref{wU}) can be
derived from Lagrangian $\mathfrak{L}=\int_{-\infty }^{+\infty }\mathcal{L}%
dx $, with density
\begin{equation}
\mathcal{L}=\frac{1}{2}U_{t}^{2}+\frac{1}{2}w_{x}^{2}+\frac{1}{2}%
U_{x}w^{2}+B_{0}U_{x}w+\frac{1}{2}R_{0}w^{2}.  \label{L}
\end{equation}%
The respective Hamiltonian is $H=\int_{-\infty }^{+\infty }\mathcal{H}dx$,
with density
\begin{equation}
\mathcal{H}=\frac{1}{2}U_{t}^{2}-\frac{1}{2}w_{x}^{2}-\frac{1}{2}%
U_{x}w^{2}-B_{0}U_{x}w-\frac{1}{2}R_{0}w^{2}.  \label{H}
\end{equation}

To define an effective potential of the interaction of two solitary waves
moving with equal velocities $v$, it is necessary to rewrite Eqs.~(\ref{U})
and (\ref{wU}), together with the Lagrangian and Hamiltonian densities (\ref%
{L}) and (\ref{H}), in the reference frame moving with velocity $v$, i.e.,
in terms of the $\tau=t$ and $\xi=x-vt$ variables, 
as:
\begin{gather}
U_{\tau\tau}-2vU_{\xi \tau}+v^{2}U_{\xi \xi }+\left( \frac{1}{2}%
w^{2}+B_{0}w\right) _{\xi }=0,  \label{Uxi} \\
w_{\xi \xi }-R_{0}w-B_{0}U_{\xi }-wU_{\xi }=0,  \label{wxi}
\end{gather}%
and%
\begin{equation}
\mathcal{L}_{\mathrm{moving}}=\frac{1}{2}U_{\tau}^{2}-vU_{\xi }U_{\tau}+%
\frac{v^{2}}{2}U_{\xi }^{2}+\frac{1}{2}w_{\xi }^{2}+B_{0}U_{\xi }w+\frac{1}{2%
}R_{0}w^{2}+\frac{1}{2}U_{\xi }w^{2},  \label{Lmoving}
\end{equation}%
\begin{equation}
\mathcal{H}_{\mathrm{moving}}=\frac{1}{2}U_{\tau}^{2}-\frac{v^{2}}{2}U_{\xi
}^{2}-\frac{1}{2}w_{\xi }^{2}-B_{0}U_{\xi }w-\frac{1}{2}R_{0}w^{2}-\frac{1}{2%
}U_{\xi }w^{2}.  \label{Hmoving}
\end{equation}

In what follows, we will introduce effective mass $M$ of the solitary wave.
In that connection, we note that, in standard models, the solitary wave's
kinetic energy, which is produced by the integral, with respect to $\xi$, of
the kinetic part of the Lagrangian density ($\int_{-\infty }^{+\infty }d\xi $%
), is \cite{kivshar}:
\begin{equation}
E_{\mathrm{kin}}=\left( 1/2\right) Mv^{2}.  \label{Ekin}
\end{equation}


\section{The interaction of far separated solitary waves}

As is customary in studies of generic settings~\cite%
{interaction,manton,panos_inter}, a pair of interacting solitary waves
separated by large distance $L$ is approximated by juxtaposing two identical
solitary-wave solutions given by Eqs.~(\ref{usol}) and (\ref{wsol}). They
interact via their exponentially decaying tails.
Assuming that the center of the solitary wave is fixed at $\xi =0$, and
taking into account that $\mathrm{sech}(x)\approx 2e^{-x}$ at $x\rightarrow
\infty $, the expressions for the tails which are derived by \eqref{wsol}
and \eqref{usol} are:
\begin{eqnarray}
w_{\mathrm{sol}} &\approx &\frac{4B_{0}}{Cv}\left( v^{2}-C^{2}\right) \exp
\left( -\frac{B_{0}}{vC}\sqrt{v^{2}-C^{2}}|\xi |\right) ,  \label{asympt-w}
\\
u_{\mathrm{sol}} &\approx &-\frac{B_{0}}{v^{2}}w_{\mathrm{sol}}=-\frac{%
4B_{0}^{2}}{Cv^{3}}\left( v^{2}-C^{2}\right) \exp \left( -\frac{B_{0}}{vC}%
\sqrt{v^{2}-C^{2}}|\xi |\right) .  \label{asympt-u}
\end{eqnarray}%
In turn, the use of Eq.~(\ref{potential}) produces the respective asymptotic
expression for the tail of field $U$:
\begin{equation}
U_{\mathrm{sol}}\approx \frac{4B_{0}}{v^{2}}\sqrt{v^{2}-C^{2}}\mathrm{sgn}%
\left( \xi \right) \exp \left( -\frac{B_{0}}{vC}\sqrt{v^{2}-C^{2}}|\xi
|\right) +c_{\pm },  \label{asympt-U}
\end{equation}%
where $c_{\pm }$ are constant values at $\xi =\pm \infty $. The average
value of the asymptotic constants, $(1/2)\left( c_{+}+c_{-}\right) $, is
arbitrary, while the difference is uniquely determined by the solution as:
\begin{equation}
c_{+}-c_{-}=\int_{-\infty }^{+\infty }u_{\mathrm{sol}}(\xi )d\xi .
\label{integral}
\end{equation}

Next, we consider the pair of solitary waves with centers placed at
positions $\xi _{0}=\pm L/2$, and the constant value of $U$ between them
[see term $c_{\pm }$ in Eq.~(\ref{asympt-U})] set equal to zero, so as to
make the configuration symmetric. Then, an effective potential of the
interaction between the far separated solitary waves, $W(L)$, can be derived
by means of the general procedure elaborated in Ref.~\cite{interaction}.
This is based on the substitution of the juxtaposition of the solitary waves
in the expression for $H$, and handling terms with spatial derivatives by
means of the integration by parts, so that the actual calculation of the
integrals is not necessary, with all the contributions from the integrals
being produced by the \textquotedblleft surface terms\textquotedblright\ in
the formula for the integration by parts. The result of this procedure is:
\begin{equation}
W(L)=\frac{32B_{0}^{3}}{C^{3}v^{3}}\left( v^{2}-C^{2}\right) ^{5/2}\exp
\left( -\frac{B_{0}}{vC}\sqrt{v^{2}-C^{2}}L\right) ,  \label{W}
\end{equation}%
with the positive sign of $W$ implying repulsion between the solitary waves.
It is relevant to mention that, when calculating the effective potential (%
\ref{W}), the result is produced by the third term in the Hamiltonian
density (\ref{Hmoving}), while the contributions from the second and fourth
ones \emph{\ exactly cancel each other}. It should be noted here that the
relation of the AA system to the KdV equation at speeds close to $C$ \cite%
{PRE}, and the pairwise repulsion of KdV solitons~\cite{kivshar} is in line
with the above analysis.

The repulsion, described above, will lead to an \emph{splitting} of the
initially equal velocities of the interacting solitary waves,
\begin{equation}
v\rightarrow v\pm \Delta v,  \label{splitting}
\end{equation}
provided that $\Delta v$ represents a small perturbative effect. To obtain $%
\Delta v$ from the energy balance, it is necessary to know the exact
expression for the energy of individual solitary waves. The substitution of
the exact solitary wave solution given by Eqs.~(\ref{usol}) and (\ref{wsol})
in the expression for the Hamiltonian, determined by its density (\ref{H}),
leads to a very cumbersome expression. This expression becomes simpler in
the limit case when the velocity is taken close to the solitary wave
existence cutoff,
\begin{equation}
v-C\ll C.  \label{<<}
\end{equation}%
Then, from Eqs. (\ref{wsol})-(\ref{xi}), we obtain
\begin{eqnarray}
u_{\mathrm{sol}} &\approx &-\frac{2B_{0}^{2}}{C^{3}}\left( v-C\right)
\mathrm{sech}^{2}\left( \frac{B_{0}\sqrt{v-C}}{\sqrt{2}C^{3/2}}\xi \right) ,
\label{uapprox} \\
w_{\mathrm{sol}} &\approx &\frac{2B_{0}}{C}\left( v-C\right) \mathrm{sech}%
^{2}\left( \frac{B_{0}\sqrt{v-C}}{\sqrt{2}C^{3/2}}\xi \right) ,
\label{vapprox} \\
H_{\mathrm{sol}} &\approx &\frac{8\sqrt{2}B_{0}^{3}}{3C^{5/2}}\left(
v-C\right) ^{3/2}.  \label{Happrox}
\end{eqnarray}%
The consideration of this case is relevant because the exponential smallness
in Eq.~(\ref{W}) is less acute for small $\left( v-C\right) $. Note, in
particular, that the $\mathrm{sech}^{2}$ limit corresponds to the soliton in
the KdV limit of the AA system \cite{PRE}.

Next, the interaction-induced change of the velocities, $\Delta v$, is
determined by equating the interaction energy (\ref{W}) to the difference
between the energy of the two-solitary-wave configuration and the sum of
individual energies of the two solitary waves, with the velocities split as
per Eq. (\ref{splitting}):
\begin{equation}
\Delta \left( H_{\mathrm{two~sol.}}\right) \approx \frac{\partial ^{2}H_{%
\mathrm{sol}}}{\partial v^{2}}\left( \Delta v\right) ^{2}\approx \frac{2%
\sqrt{2}B_{0}^{3}}{C^{5/2}\sqrt{v-C}}\left( \Delta v\right) ^{2},
\label{Delta}
\end{equation}%
where condition (\ref{<<}) is used to simplify the expression, as it follows
from Eq.~(\ref{Happrox}). Finally, equation $\Delta \left( H_{\mathrm{%
two~sol.}}\right) =W(L)$ yields the result, which is valid under condition (%
\ref{<<}), provided that the result also satisfies the constraint $\Delta
v\ll v-C$ (i.e., it is a small perturbative effect):
\begin{equation}
\Delta v\approx \frac{8}{\sqrt{C}}\left( v-C\right) ^{3/2}\exp \left( -\frac{%
B_{0}\sqrt{v-C}}{\sqrt{2}C^{3/2}}L\right) .  \label{Delta-v}
\end{equation}

Note that, for fixed large $L$ and fixed $C$ and $B_{0}$, the
interaction-induced velocity change, $\Delta v$, as given by Eq. (\ref%
{Delta-v}) and considered as a function of $\left( v-C\right) $, attains a
maximum (the strongest perturbative effect of the interaction) at
\begin{equation}
(v-C)|_{\max }=18C^{3}/\left( B_{0}L\right) ^{2},  \label{v-C}
\end{equation}%
the maximum value itself being
\begin{equation}
\left( \Delta v\right) _{\max }=\left( \frac{6\sqrt{2}}{e}\right) ^{3}\frac{%
C^{4}}{\left( B_{0}L\right) ^{3}}\approx 30\frac{C^{4}}{\left( B_{0}L\right)
^{3}}.  \label{max}
\end{equation}%
The above calculation, albeit approximate (especially since the velocity
difference will keep changing as separation $L$ changes), suggests an
important qualitative observation that is corroborated below by numerical
computations. In particular, if we start from a symmetric configuration, it
will progressively become asymmetric, leading to a pattern with taller and
faster solitary waves on the right, and a shorter, slower solitary waves on
the left. Below it is confirmed that, indeed, such a configuration in terms
of heights and speeds is formed in the case of multi-solitary wave states.

In a quantitative form, the relative motion of interacting solitary waves
obeys the dynamical equation,
\begin{equation}
\frac{d^{2}L}{d\tau ^{2}}=-\frac{1}{M_{\mathrm{reduced}}}\frac{dW}{dL}.
\label{dtnamics}
\end{equation}%
Here, the reduced mass $M_{\mathrm{reduced}}$ of the solitary wave pair is
considered to be
\begin{equation}
M_{\mathrm{reduced}}=(1/2)M,  \label{M/2}
\end{equation}%
where we have adopted the particle-like nature of the solitary wave,
together with the standard result of the classical mechanics concerning the
interaction of two point masses; thus, in Eq.~(\ref{M/2}), $M$ represents
the above-mentioned mass of a single solitary wave, see Eq.~(\ref{Ekin}).
Actually, the calculation of $M$ is a central point in our analysis, as it
concerns the comparison with numerical results (see next Section and
Appendix~\ref{appendix_force}). We also note that the above approach to the
the prediction of the evolution of the separation between the solitary waves
is based on the energetics of the multi-solitary wave ansatz. A systematic
analysis, including all the associated technical details at the level of the
PDE system and conservation laws, generalizing another approach, introduced
by Manton~\cite{manton}, is presented in Appendix~A. We demonstrate that it
leads to the same result as Eq.~(\ref{dtnamics}), thus confirming the above
findings.

\section{Numerical results}

\subsection{The simulations}

To perform a numerical study of the dynamics of interacting solitary waves
in the AA model, we first consider a pair of solitary waves with equal
velocities,
\begin{equation}
v_{1,2}(t=0)=1.1,  \label{1.1}
\end{equation}
which are initially placed at $x_{1,2}(t=0)=\pm 10$, i.e.,~the initial
distance between them is $L(0)=x_{1}(t=0)-x_{2}(t=0)=20$. In all
simulations, we set $B_{0}=R_{0}=1$ (for this choice, the dimensionless form
of the AA model considered here coincides with the one adopted in Ref.~\cite%
{nairn}), which means $C=1$, see Eqs.~(\ref{R})-(\ref{R0}).
To apply the numerical method, the spatial variable was discretized by
finite differences, and forward marching in time was performed.
The finite-difference scheme in space was implemented taking into regard the
coupling of a given site with second neighbors, in order to produce a stable
numerical algorithm. The time-integration has been performed with the
Runge-Kutta method of the $4^{\mathrm{th}}$-$5^{\mathrm{th}}$ order. To
check the precision of the results, we used, as a diagnostic, the value of
the total energy of the system calculated according to Eq. \eqref{H}.
Relative variance of the energy in all the simulations was $<10^{-6}$.

Since the physically relevant interval of the solitary-waves' velocities is,
according to (\ref{C}), $1<v<2$, the selected value (\ref{1.1}) is close to
the lower edge of the interval. As can be inferred from Eqs.~\eqref{usol}-%
\eqref{wsol}, the solitary waves in this velocity region are wide, which
implies that their interaction is stronger; as a result, shorter integration
times are required to let the interaction manifest itself.

\begin{figure}[h]
\includegraphics[scale=0.45]{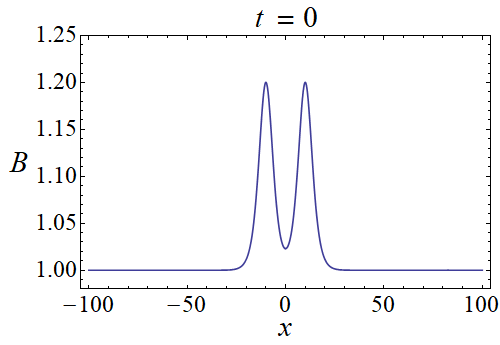}\hspace{0.5cm}%
\includegraphics[scale=0.45]{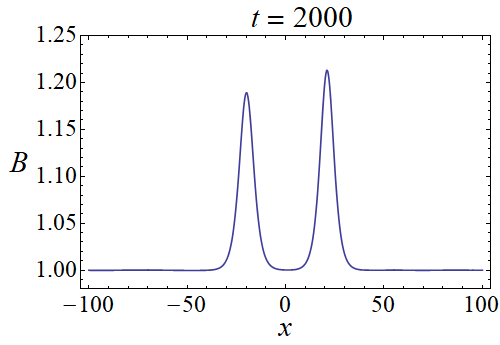}
\caption{Results of simulations of the interaction of two solitary waves, as
they are observed in the co-travelling reference frame for initial
velocities $v_{1,2}(t=0)=1.1$ and positions $x_{1,2}(t=0)=\pm 10$. The two
solitary waves, which initially have equal velocities, end up with different
ones and, consequently, unequal heights.}
\label{fig:TwoSolitons_v_1_1_x_10}
\end{figure}

The dynamical behavior of the two interacting solitary waves, in the
co-travelling reference frame, is shown in Fig.~\ref%
{fig:TwoSolitons_v_1_1_x_10}. The symmetric configuration is quickly
converted into one in which one solitary wave becomes taller (and
consequently quicker) than the other. This is expected, as predicted by the
analysis of the previous Section, since the repelling interaction of the
solitary waves causes a change of their velocities, $\Delta v_{1}>0$ and $%
-\Delta v_{2}<0$, so that the velocities resulting from the interaction are
\begin{equation}
v_{1}=v_{1}(t=0)+\Delta v_{1}>v_{2}=v_{2}(t=0)-\Delta v_{2}.  \label{v1v2}
\end{equation}%
The situation is reversed when we consider initial velocities with the
opposite sign; in this case, the resulting configuration is a mirror image
of Fig.~\ref{fig:TwoSolitons_v_1_1_x_10} (not shown here).


%
%

\begin{figure}[h]
\includegraphics[scale=0.45]{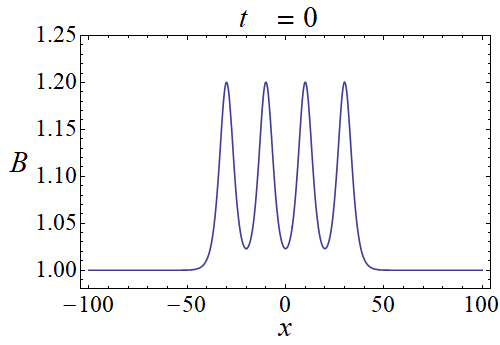}\hspace{0.5cm}%
\includegraphics[scale=0.45]{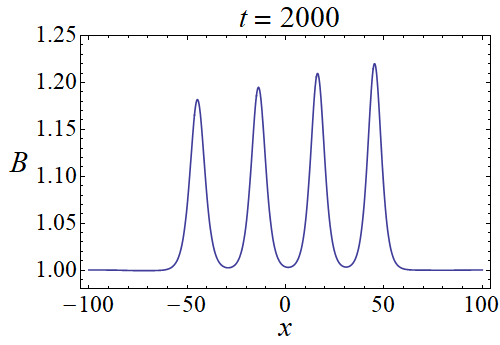}
\caption{The result of the simulation of the set of four interacting
solitary waves in the co-travelling reference frame. The initial velocities
are $v_{1,2,3,4}(0)=1.1$, and their initial positions are $x_{1,4}(0)=\pm 30$
and $x_{2,3}(0)=\pm 10$. The solitary waves end up with different velocities
and, consequently, different heights, cf. Fig.~\protect\ref%
{fig:TwoSolitons_v_1_1_x_10}.}
\label{fig:daltons}
\end{figure}

Similar phenomenology is observed in the simulations if we consider more
than two solitary waves placed symmetrically, with equal initial velocities.
In particular, in Fig.~\ref{fig:daltons} we display a four-solitary-wave
configuration. In this case, the result shows %
%
a graded configuration of increasingly taller and faster solitary waves,
that keep separating from each other in the course of the subsequent
evolution. 

In addition, we have considered the evolution of solitary-wave sets with
higher initial velocities (equal for all the pulses in the set, and with the
same initial distance between them), taken in the range of $1.25\leqslant
v(t=0)\leqslant 1.9$. We observed the same phenomenology but, as the width
of the waves becomes smaller with the increase of $v$, the interaction
becomes, accordingly, weaker and the corresponding dynamical response is
slower than in the above case of $v(t=0)=1.1$.

\subsection{Comparison of the analytical estimate with numerical simulations}

From the predicted form of the interaction potential \eqref{W} and
expression (\ref{M/2}) for the reduced mass, we derive the equation of
motion for the distance between the two solitary waves:
\begin{equation}
\frac{d^{2}L}{d\tau ^{2}}=-\frac{2}{M}\frac{d}{dL}W(L)=\frac{2A(v)}{M}\exp
\left( -\lambda L\right) ,  \label{eq:eq_motion}
\end{equation}%
where
\begin{equation}
A(v)\equiv \frac{32B_{0}^{4}}{C^{4}v^{4}}\left( v^{2}-C^{2}\right)
^{3},\quad \lambda \equiv \frac{B_{0}}{vC}\sqrt{v^{2}-C^{2}}.~
\label{Alambda}
\end{equation}%
A subtle issue in this connection is the identification of the mass $M$ of
the solitary wave. Considering either the kinetic energy term in the
Hamiltonian (for a stationary solitary wave in the co-traveling reference
frame) and setting it equal to $(1/2)Mv^{2}$, or the momentum, 
\begin{equation}
P=-\int_{-\infty }^{+\infty }U_{t}U_{x}dx,  \label{P}
\end{equation}%
and setting it equal to $Mv$, leads to the conclusion that
\begin{equation}
M=\int_{-\infty }^{+\infty }u^{2}dx.  \label{M}
\end{equation}%
In Appendix~\ref{appendix_force} we further
explore this definition of the solitary wave's mass, upon considering the
dynamical response of the solitary wave to a perturbation represented by a
potential term added to the system, which is also corroborated by direct
numerical simulations.

Here it should be pointed out that the above expression for $M$ is not an
intuitively evident one, as it refers solely to the mass associated with the $%
u$-component of the AA solitary wave, while the $w$-component does not
contribute to the calculation of the mass, because Eq. (\ref{w}) for this
component does not contain time derivatives. In light of this fact, here we
proceed in the following way: by numerically solving the ordinary
differential equation (ODE)~\eqref{eq:eq_motion}, we obtain distance $L$
between the two solitary waves as a function of time. This prediction is
compared to the full numerical result, produced by simulations of the AA
system, i.e., Eqs. (\ref{u}) and (\ref{w}). Assuming that the tail-tail
interaction force, produced by both the energetic considerations and the
Manton method (see Appendix~A) adequately characterizes the exponential
nature of the pairwise repulsion between the solitary waves, we then use the
above semi-analytical prediction and its comparison to the full numerical
results to \textquotedblleft adjust\textquotedblright\ the proper expression
for the mass. This approach reveals a relevant correction to the effective
mass of the solitary wave.

As said above, we aim, first, to numerically integrate Eqs.~\eqref{u} and %
\eqref{w} and thus obtain the distance between the interacting solitary
waves as a function of time. For this purpose, we used velocities $%
v\geqslant 1.5$, to make the waves more well-separated and thus improve the
accuracy of the comparison of the full numerical results with predictions of
the ODE (\ref{eq:eq_motion}), where the constants and the \textquotedblleft
naively defined" solitary-wave's mass are taken as per Eqs.~(\ref{Alambda})
and (\ref{M}), respectively. The results for $v=1.6$ are shown in Fig.~\ref%
{fig:TwoSolitons_v_1_6_x_10_t_100_comparison}, where the red solid and
dashed blue lines show, respectively, the distance between the solitary
waves, as obtained from the direct numerical integration of Eqs.~\eqref{u}
and \eqref{w}, and predicted by the solution of the ODE~\eqref{eq:eq_motion}%
. In this case, the discrepancy between the PDE and ODE results is obvious.
Similar results are produced by the comparison at other values of the
parameters.

\begin{figure}[tbp]
\includegraphics[scale=0.35]{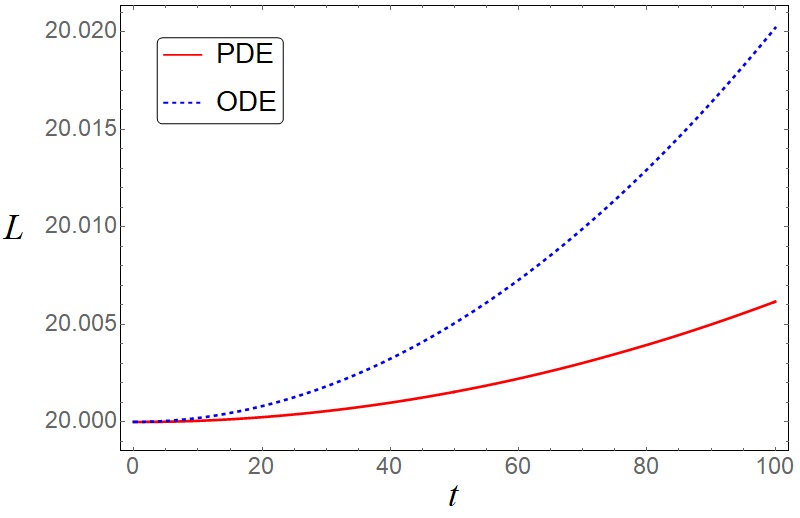}
\caption{The comparison of the results for distance $L$ between the two
interacting solitary waves in the co-travelling reference frame, for initial
velocities $v_{1,2}(0)=1.6$. The (red) solid and (blue) dashed lines present
the results produced, severally, by direct simulations of the underlying
PDEs~(\protect\ref{u}) and (\protect\ref{w}), and by the numerical solution
of the effective ODE~\eqref{eq:eq_motion} with the constants taken as per
Eq.~(\protect\ref{Alambda}) and (\protect\ref{M}).}
\label{fig:TwoSolitons_v_1_6_x_10_t_100_comparison}
\end{figure}

%

Following the path outlined above, we attribute the discrepancy to the
uncertainty regarding the solitary-wave's mass. To fix the issue, we
\textquotedblleft phenomenologically" incorporate a fitting factor $\alpha
(v)$ in Eq.~\eqref{eq:eq_motion}, rewriting it as
\begin{equation}
\frac{d^{2}L}{d\tau^{2}}=\frac{A(v)}{M_{\mathrm{reduced}}^{\ast }}\exp
\left( -\lambda L\right) ,\ \ \mathrm{with}\quad M_{\mathrm{reduced}}^{\ast
}=\frac{M}{2\alpha (v)}.  \label{eq:eq_motion3}
\end{equation}%
Then, we determine the value of $\alpha (v)$ required for the PDE- and
ODE-produced curves to match. The results are shown, for $v=1.6$ and other
values of the initial velocities, in Fig.~\ref%
{fig:TwoSolitons_v_1_6_x_10_t_100_comparison_fixed} and Table~\ref{tab:a_v}.
In particular, for $v=1.6$ the two curves are made virtually identical by
dint of the adjustment factor $\alpha =0.3045$ in Eq.~(\ref{eq:eq_motion3}).
This observation and similar findings for other initial velocities confirm
that the above analysis correctly captures the exponential decay of the
inter-solitary-wave repulsive force, yet the straightforward theory misses
the right prefactor in the respective equation of motion (\ref{eq:eq_motion}%
). 
In this connection, we stress that, for our analytis to be relevant for the
comparison to the full simulations, we need the solitary waves to be well
separated, in order apply the assumption of the tail-to-tail interaction,
but not too far from each other either, lest the integration time, needed to
make the interaction effect tangible, should be extremely large. In the
present setting, we considered the initial separation of $L(0)=20$, and we
did not observe any significant difference in the accuracy of the obtained
results for larger separations. 
%
\begin{figure}[tbp]
\includegraphics[scale=0.29]{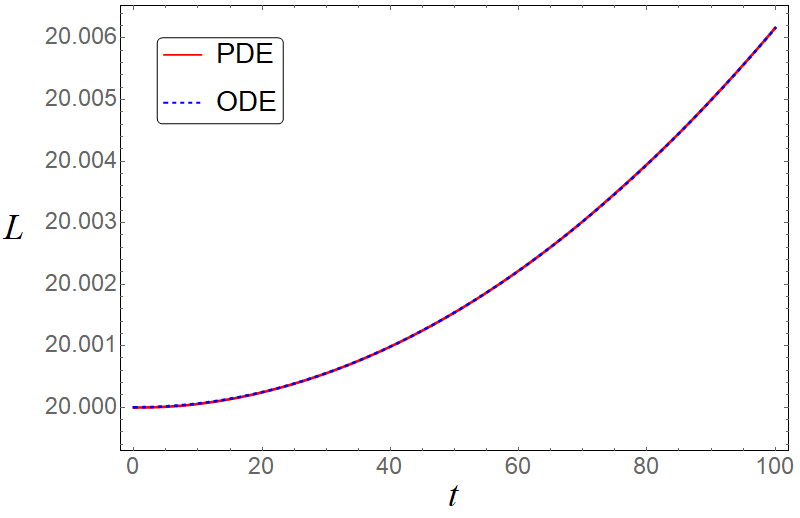}
\includegraphics[scale=0.28]{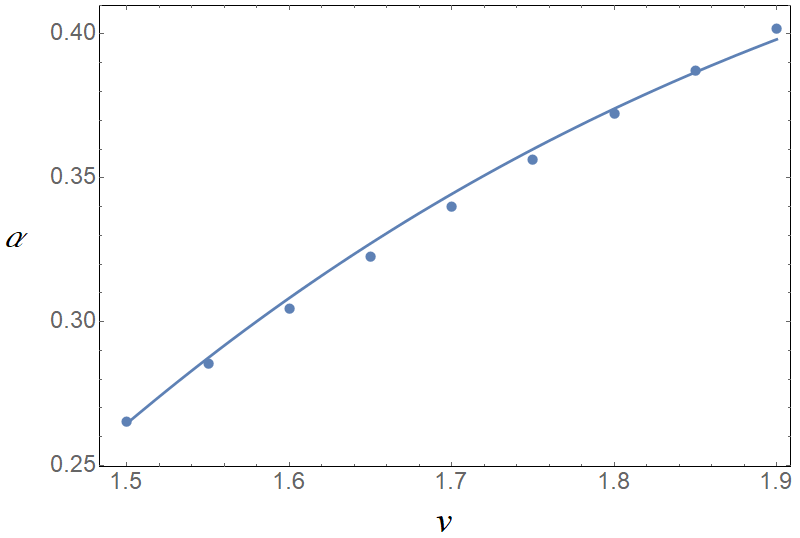}
\caption{The left panel shows the comparison between the PDE results [solid
(red) curve], the same as in Fig.~\protect\ref%
{fig:TwoSolitons_v_1_6_x_10_t_100_comparison}), for distance $L(t)$ between
the interacting solitary waves in the co-travelling reference frame, and the
ODE counterpart [dashed (blue) curve], produced by Eq.~(\protect\ref%
{eq:eq_motion3}) using the fitting prefactor $\protect\alpha =0.3045$, in
the case of $v_{1,2}(0)=1.6$. It is seen that this value of $2\protect\alpha
$ makes the two curves practically identical. The right panel shows, by
means of the chain of dots, the fitting half-factor $\protect\alpha $ for
different values of the initial speed, as per the data presented in Table~%
\protect\ref{tab:a_v}. The continuous curve plots an interpolating function (%
\protect\ref{alpha}), which approximates the values of $\protect\alpha (v)$.}
\label{fig:TwoSolitons_v_1_6_x_10_t_100_comparison_fixed}
\end{figure}
%
%
\begin{table}[h]
\begin{tabular}{|c|c|c|c|c|c|c|c|c|c|}
\hline
$v$ & 1.5 & 1.55 & 1.6 & 1.65 & 1.7 & 1.75 & 1.8 & 1.85 & 1.9 \\ \hline
$\alpha$ & 0.2653 & 0.2855 & 0.3045 & 0.3227 & 0.34 & 0.3565 & 0.3724 &
0.3874 & 0.4019 \\ \hline
\end{tabular}%
\caption{Values of the fitting half-factor $\protect\alpha $ in Eq. (\protect
\ref{eq:eq_motion3}) which provide the best match of $L(t)$ to the results
of PDE simulations at different values of $v$.}
\label{tab:a_v}
\end{table}

To summarize these results, we sought a function $\alpha =\alpha (v)$ which
may fit the data from Table \ref{tab:a_v}. As it is seen in the right panel
of Fig.~\ref{fig:TwoSolitons_v_1_6_x_10_t_100_comparison_fixed},
a reasonable choice is
\begin{equation}
\alpha (v)=1.05(v^{2}-C^{2})^{2}/v^{9/2}.  \label{alpha}
\end{equation}%
This function is built as a combination of powers of $(v^{2}-C^{2})$ and $v$%
, as these factors naturally appear in the calculation of the interaction
force for the solitary waves. In Appendix~\ref{appendix_force}, we evaluate
the relevance of considering the variation of the solitary wave mass in the
context of the solitary wave-defect interaction.


\section{Conclusions}

In the present work, we have revisited the Adlam-Allen (AA) model, governing
the propagation of solitary waves in cold magnetized collisionless plasmas,
in the presence of the electric field (in addition to an magnetic field).
The AA model is one of the fundamental nonlinear models of plasma physics~%
\cite{aa1,aa2,allen3,allen4} that has made the prediction of solitary waves
possible, well before the (re-)discovery of the KdV equation and its
celebrated solitons in the framework of the Fermi-Pasta-Ulam-Tsingou model.
Indeed, the AA system is a source of localized and periodic waves, not only
in the context of the transverse magnetic field applied to the plasmas, but
also more recently for a longitudinal field~\cite{gohar1,gohar2}.

Here, we have studied the interaction between solitary waves, a theme of
substantial interest in the theory of solitary waves and solitons \cite%
{manton,kivshar,interaction}. We provide the energy analysis, based on the
Hamiltonian structure of the model, and complement it with a detailed
derivation of the same result by means of an alternative (Manton's) method,
see Appendix. The resulting Newtonian dynamics for the separation clearly
reveals the repulsive character of the interaction, as well as the
exponential dependence of the force on the separation. This is natural to
expect near the lower edge of the range (\ref{C}) of accessible solitary
wave speeds, where the model is close to the KdV limit (as shown earlier in
Refs.~\cite{PRE,nairn}) and, thus, inherits the repulsive interaction
between solitons which is well known in the framework of the KdV equation.
Nevertheless, an essential element of uncertainty remains in the form of an
accurate expression for the effective mass of the two-component solitary
wave. We have side-stepped this uncertainty by finding a suitable
velocity-dependent fitting factor, which takes values, roughly, between $%
2\alpha =0.5$ and $0.8$, which depends on the wave's speed. This prefactor
secures the full match between the ODE (semi-analytical) and PDE (fully
numerical) results for the separation between the interacting solitary waves.

While our analysis provides a definitive explanation for the exponentially
repulsive nature of the interactions, a remaining intriguing issue concerns
the speed-dependent prefactor in the respective effective equation of motion
fot the separation between the interacting solions. This amounts to an
effective renormalization of the solitary-wave's dynamical mass. The same
issue also concerns the dynamics of the soliton gas in the AA system. A
prototypical example of the latter was demonstrated in Fig.~\ref{fig:daltons}
for a configuration consisting of four interacting solitary waves, initially
having equal velocities, which end up with different velocities and,
consequently, different heights. 
This issue may also be relevant for other effectively nonlocal systems, in
which one equation does not contain time derivatives (e.g., a Poisson-like
equation). Systems of the latter type arise, in particular, in models of
thermal media, plasmas, nematic liquid crystals, and Bose-Einstein
condensates, see recent examples in~works \cite{djf,ECNU} and references
therein. Such systems, as well as higher-dimensional plasma models are
natural objects for future work. Progress along these directions will be
reported elsewhere.

\section*{Acknowledgments}

The work of B.A.M. was supported, in part, by the Israel Science Foundation
through grant No. 1286/17. This material is based upon work supported by the
US National Science Foundation under Grant DMS-1809074 (P.G.K.).
Constructive discussion with Y. Kominis and I. Kourakis are gratefully
acknowledged.

\appendix

\section{Derivation of the potential of the solitary waves interaction via
Manton's approach}

Here, we aim to derive the interaction potential \eqref{W} by means of
another approach, namely upon following the Manton's method \cite{manton}.

%
Introducing the traveling coordinate $\xi $, as per Eq.~(\ref{xi}), in Eqs. %
\eqref{U}-\eqref{wU} we obtain:
\begin{gather}
U_{\tau\tau}-2vU_{\xi \tau}+v^{2}U_{\xi \xi }+B_{0}w_{\xi }+\frac{1}{2}%
(w^{2})_{\xi }=0  \label{eq:eqs44} \\
w_{\xi \xi }-R_{0}w-B_{0}U_{\xi }-U_{\xi }w=0.
\end{gather}%
%
%
%
%
In the context of Klein-Gordon equations, the Manton's method explores the
evolution of momentum $P$ (as its time derivative is associated with the
force, which here stems solely from the inter-solitary-wave interaction)~%
\cite{manton}. The expression for the field mommentum in the co-moving frame
is given by
\begin{equation}
\tilde{P}=\int_{-\infty }^{+\infty } U_{\tau}U_{\xi } d\xi.
\end{equation}
Differentiating the above expression momentum in time, we obtain:
\begin{equation}
\begin{array}{rcl}
\displaystyle\frac{d\tilde{P}}{d\tau} & = & -\displaystyle\int_{-\infty
}^{+\infty }\left( U_{\tau\tau}U_{\xi }+U_{\tau}U_{\xi \tau}\right)d\xi \\%
[12pt]
& = & -\displaystyle\int_{-\infty }^{+\infty }\left[\left(2vU_{\xi
\tau}-v^{2}U_{\xi \xi }-B_{0}w_{\xi }-\frac{1}{2}(w^{2})_{\xi }\right)
U_{\xi }+U_{\tau}U_{\xi \tau}\right] d\xi \\[12pt]
& = & \displaystyle\int_{-\infty }^{+\infty }\left[ (B_{0}+w)w_{\xi }U_{\xi
}~-\frac{1}{2}(U_{\tau}^{2})_{\xi }~{-v(U_{\xi }^{2})_{\tau}~+\frac{1}{2}%
v^{2}(U_{\xi }^{2})_{\xi }}\right] {d\xi }.%
\end{array}
\label{eq:dpdtU}
\end{equation}%
%
%

In line with the original approach of~\cite{manton}, we proceed by
considering two well-separated solitary waves, one placed at $x=0$ and the
other one at $x=L\gg 0$. We also set two points, $x=a$ and $x=b$, with $%
a\rightarrow -\infty $ and $0\ll b\ll L$. Next, following Ref. \cite{manton}%
, we neglect the two middle terms in the above equation, as we consider
quasi-stationary solutions, and by using \eqref{U} we get:
\begin{equation}
\frac{d\tilde{P}}{dt}=-\int_{a}^{b}\left[ (R_{0}w-w_{\xi \xi })w_{\xi }-%
\frac{v^{2}}{2}(u^{2})_{\xi }~\right] d\xi =\left[ \frac{1}{2}w_{\xi }^{2}-%
\frac{1}{2}R_{0}w^{2}{+\frac{v^{2}}{2}u^{2}}\right] _{a}^{b}.
\label{eq:dpdt}
\end{equation}%
Now, we consider the superposition ansatz for the far separated solitary
waves, $w=w_{1}+w_{2}$. We use the fact that the decay of the tails of the
solitary waves is exponential. Thus, the contributions at $x=a$ vanish,
while those at $x=b$ are of the form:
\begin{equation}
w_{1}\sim e^{-\lambda \xi }\quad \mathrm{and}\quad w_{2}\sim e^{\lambda (\xi
-L)}.
\end{equation}%
Then, the contributions in \eqref{eq:dpdt} which are mixed (and consequently
account for the interaction) are at point $x=b$,
\begin{equation}
\frac{d\tilde{P}}{d\tau}=w_{1_{\xi }}w_{2_{\xi }}-{R_{0}}w_{1}w_{2}+{%
v^{2}u_{1}u_{2}}.  \label{eq:dpdt_int}
\end{equation}%
The asymptotic form \eqref{asympt-w} and \eqref{asympt-u} yields
\begin{equation}
\left.
\begin{array}{rcl}
w_{1} & \approx & \displaystyle\frac{4B_{0}}{Cv}(v^{2}-C^{2})e^{-\lambda \xi
} \\[10pt]
w_{2} & \approx & \displaystyle\frac{4B_{0}}{Cv}(v^{2}-C^{2})e^{\lambda (\xi
-L)}%
\end{array}%
\right\} \Rightarrow
\begin{array}{rcl}
w_{1{\xi }} & \approx & \displaystyle-\frac{4B_{0}^{2}}{C^{2}v^{2}}%
(v^{2}-C^{2})^{3/2}~e^{-\lambda \xi }, \\[10pt]
w_{2{\xi }} & \approx & \displaystyle\frac{4B_{0}^{2}}{C^{2}v^{2}}%
(v^{2}-C^{2})^{3/2}~e^{\lambda (\xi -L)},%
\end{array}%
\end{equation}%
and
\begin{equation}
\begin{array}{rcl}
u_{1} & \approx & \displaystyle-\frac{4B_{0}^{2}}{Cv^{3}}(v^{2}-C^{2})e^{-%
\lambda \xi }, \\[10pt]
u_{2} & \approx & \displaystyle-\frac{4B_{0}^{2}}{Cv^{3}}(v^{2}-C^{2})e^{%
\lambda (\xi -L)},%
\end{array}%
\end{equation}%
with $\displaystyle\lambda =\left( B_{0}/Cv\right) \sqrt{v^{2}-C^{2}}$.
Thus, by also using $\displaystyle R_{0}=B_{0}^{2}/C^{2}$, Eq. %
\eqref{eq:dpdt_int} reads
\begin{equation}
\begin{array}{rl}
\displaystyle\frac{d\tilde{P}}{d\tau} & =-\displaystyle\frac{16B_{0}^{4}}{%
C^{4}v^{4}}(v^{2}-C^{2})^{3}~e^{-\lambda L}-\frac{16B_{0}^{4}}{C^{4}v^{2}}%
(v^{2}-C^{2})^{2}~e^{-\lambda L}{+\frac{16B_{0}^{4}}{C^{2}v^{4}}%
(v^{2}-C^{2})^{2}~e^{-\lambda L},}%
\end{array}%
\end{equation}%
or
\begin{equation}
\frac{d\tilde{P}}{d\tau}=-\displaystyle{\frac{32B_{0}^{4}}{C^{4}v^{4}}%
(v^{2}-C^{2})^{3}}~e^{-\lambda L}.
\end{equation}

Then, according to Eq.~(\ref{dtnamics}), we get
\begin{equation}
\frac{d^{2}L}{d\tau^{2}}=\frac{1}{M_{\mathrm{reduced}}}\displaystyle\frac{%
32B_{0}^{4}}{C^{4}v^{4}}(v^{2}-C^{2})^{3}~e^{-\lambda L}.  \label{ode_panos}
\end{equation}%
For the potential, we get
\begin{equation}
W(L)=\displaystyle\frac{32B_{0}^{3}}{C^{3}v^{3}}(v^{2}-C^{2})^{5/2}~e^{-%
\lambda L}.
\end{equation}%
These expressions recover the results that were obtained in the main text
via the energetic arguments [Eq.~(\ref{W})], and thus corroborate the
repulsive exponentially decaying interaction between the solitary waves,
mediated by their tails.

\section{Interaction of the solitary wave with a parametric force}

\label{appendix_force} Our aim in the present Appendix is to study the
interaction between a solitary wave of the system with a defect using
Manton's approach and to assess the relevance of mass variation during such
an interaction. To do so, we introduce a perturbation term $F$ in Eq.~(\ref%
{u}), as follows:
\begin{equation}
u_{tt}+\left( \frac{1}{2}w^{2}+B_{0}w\right) _{xx}=F,\ \ \mathrm{with}\quad
F\equiv f(x)u(x,t).  \label{F}
\end{equation}%
The reasoning behind this choice of perturbation and the form of $f(x)$
is the following. Arguably, one of the simplest possibilities is to define a
spatially localized parametric drive which introduces a small localized
perturbation to the motion of the solitary wave, and is compatible with the
zero boundary conditions as $x\rightarrow \pm \infty $ (in fact, as $%
x\rightarrow \pm \tilde{L}/2$, where $\tilde{L}$ is the size of the
integration domain). 
%
%
%
%
In particular, we select
\begin{equation}
f\equiv f_{0}\frac{\sinh \left( x/l\right) }{\cosh ^{3}\left( x/l\right) }.
\label{f(x)}
\end{equation}%
Here $f_{0}$ and $l$ characterize, respectively, the amplitude and width of
the spatially localized parametric perturbation term. In the subsequent
numerical investigation, we considered a solitary wave with velocity $%
v_{0}=1.7$, under the action of the perturbation 
of the above form, with $f_{0}=0.1$ and $l=1.5$.

\begin{figure}[h]
\includegraphics[scale=0.4]{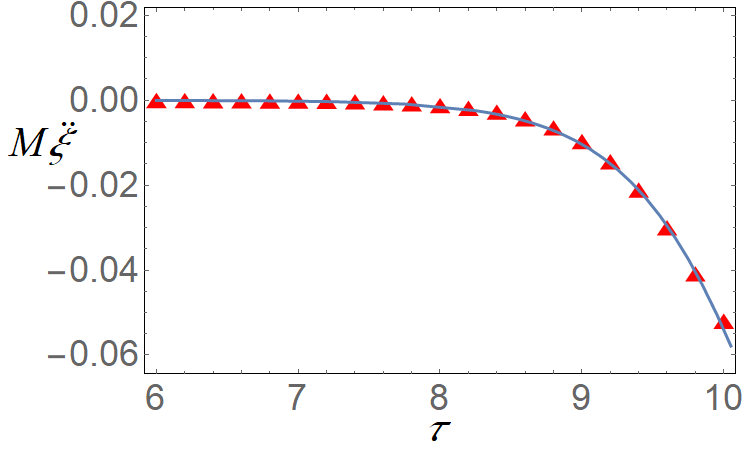}
\caption{Comparison of the right-hand sides of Eqs.~\eqref{eq1} and
\eqref{eq2}. The blue solid line and chain oif red triangles depict,
respectively, the right-hand-sides of Eqs. \eqref{eq1} and \eqref{eq2}.}
\label{fig:comparison1}
\end{figure}

To utilize the Manton's approach, we use the system's equation of motion
written in terms of field $w$, potential $U$ and the co-traveling
coordinates $(\xi ,\tau )$. This way the system becomes

\begin{equation}
U_{\tau \tau }-2vU_{\xi \tau }+v^{2}U_{\xi \xi }+B_{0}w_{\xi }+\frac{1}{2}%
(w^{2})_{\xi }=\int_{-\infty }^{\xi }f\frac{\partial U}{\partial \xi
^{\prime }}\,d\xi ^{\prime }.  \label{c1}
\end{equation}%
We recall that the momentum of the soliton in the moving frame is given by
\begin{equation}
\tilde{P}=-\int_{-\infty }^{\infty }U_{\tau }U_{\xi }\,d\xi ,  \label{c2}
\end{equation}%
and its derivative by
\begin{equation}
\frac{d\tilde{P}}{d\tau }=-\int_{-\infty }^{+\infty }U_{\tau \tau }U_{\xi
}\,d\xi -\int_{-\infty }^{+\infty }U_{\tau }U_{\xi \tau }\,d\xi .  \label{c3}
\end{equation}%
By substituting the $U_{\tau \tau }$ term from Eq. \eqref{c1} into Eq. %
\eqref{c3}, we obtain
\begin{equation}
\begin{array}{rl}
\displaystyle\frac{d\tilde{P}}{d\tau }= & \displaystyle-2v\int_{-\infty
}^{+\infty }U_{\xi \tau }U_{\xi }d\xi +v^{2}\int_{-\infty }^{\infty }U_{\xi
\xi }U_{\xi }\,d\xi +\int_{-\infty }^{+\infty }\left[ B_{0}w_{\xi }+\frac{1}{%
2}(w^{2})_{\xi }\right] U_{\xi }\,d\xi  \\[15pt]
& \displaystyle-\int_{-\infty }^{+\infty }U_{\xi }\int^{\xi }f\frac{\partial
U}{\partial \xi ^{\prime }}\,d\xi ^{\prime }\,d\xi -\int_{-\infty }^{+\infty
}U_{\tau }U_{\xi \tau }\,d\xi .%
\end{array}
\label{eq1}
\end{equation}%
On the other hand, we assume that $U=U(\xi -\xi _{0}(\tau ))$, which
represents a traveling-wave solution with its (moving) center $\xi _{0}(\tau
)$. This is the only essential assumption adopted for the current analysis,
i.e., we are exploring the dynamics of an adiabatically varying solitary
wave whose properties are slowly varying in the presence of the defect.
Thus, we set $U_{\tau }=-U_{\xi }\,\dot{\xi _{0}}$ and from Eq. \eqref{c2}
we get, also using definition \eqref{M},
\begin{equation}
\tilde{P}=\dot{\xi _{0}}\int_{-\infty }^{\infty }U_{\xi }^{2}\,d\xi =\dot{%
\xi _{0}}\int_{-\infty }^{\infty }u^{2}\,d\xi =\dot{\xi _{0}}\,M
\end{equation}%
and, accordingly,
\begin{equation}
\frac{d\tilde{P}}{d\tau }=M\,\ddot{\xi _{0}}+\dot{M}\dot{\xi _{0}}
\label{eq2}
\end{equation}%
The right-hand sides of Eqs. (\ref{eq1}) and (\ref{eq2}) should match.
Indeed, if we consider the numerical solution of the system for $U$, $w$ to
the calculate mass $M$, we see in Fig.~\ref{fig:comparison1} that they
match. Note that we only consider the evolution up to $\tau =10$, rather
than the full time interval of the interaction between the solitary wave and
the potential. This is done because when the solitary wave interacts fully
with the defect, it is deformed (as it can be seen in Fig.~\ref%
{fig:SingleSoliton_forced_v_1_7} for $\tau =16$) and the
adiabatic-traveling-wave assumption is no longer valid, broken by the
emission of radiation, as observed in the bottom panels of Fig.~\ref%
{fig:SingleSoliton_forced_v_1_7}. On the other hand, up to $\tau =10$ it is
relevant to assume that only the tail of the solitary wave and defect's
potential interact, the shape of the wave being only adiabatically modified,
in line with the underlying assumptions. In Figure~\ref%
{fig:SingleSoliton_forced_v_1_7}, we consider the evolution in the co-moving
frame, thus the main body of the solitary wave appears still, while the
inhomogeneity moves with negative velocity $v^{\prime }=-1.7$.

Let us now evaluate terms contributing to the change of the momentum in the
force balance associated with this problem. The contribution of the $\dot{M}%
\dot{\xi _{0}}$ term in Eq. (\ref{eq2}) turns out to be negligible in
comparison with the $M\ddot{\xi _{0}}$ one, being at least two orders of
magnitude smaller than the latter one. For example, for $\tau ^{\ast }=9.5$
we have $M(\tau ^{\ast })\ddot{\xi _{0}}(\tau ^{\ast })=-0.0251229$ and $%
\dot{\xi _{0}}(\tau ^{\ast })\dot{M}(\tau ^{\ast })=0.000109553$.
Notice that in this setting the mass is evaluated to be: $M(\tau ^{\ast
})=1.7148755$, the acceleration is $\ddot{\xi}(\tau ^{\ast })=-0.01465$,
while the arising change in the mass is characterized by $\dot{M}(\tau
^{\ast })=-0.0147101$ and $\dot{\xi}(\tau ^{\ast })=-0.0251229$. It is thus
concluded that, in the course of the time interval where this Manton-type
method force balance is applicable, the variation of the mass is explicitly
estimated to be $\approx 1\%$ in comparison to the actual solitary-wave
mass, hence, for the time interval of our current considerations, the mass
change does not substantially affect the wave-defect interaction.

\begin{figure}[h]
\includegraphics[scale=0.25]{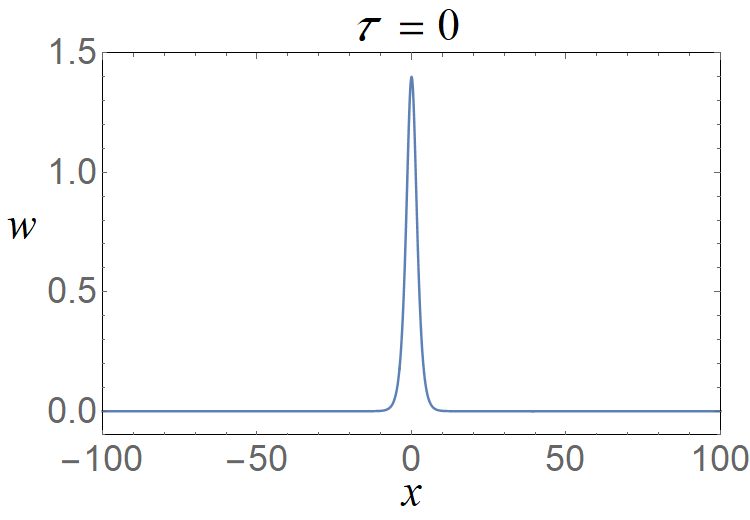}%
\hspace{0.3cm}%
\includegraphics[scale=0.25]{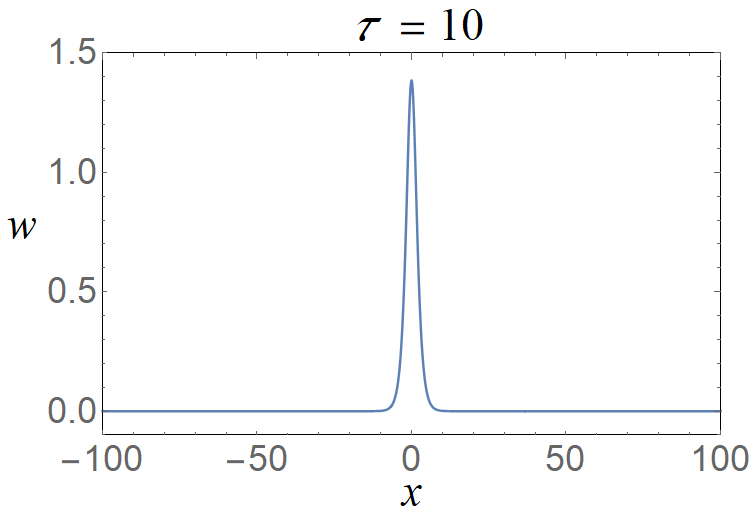} %
\includegraphics[scale=0.25]{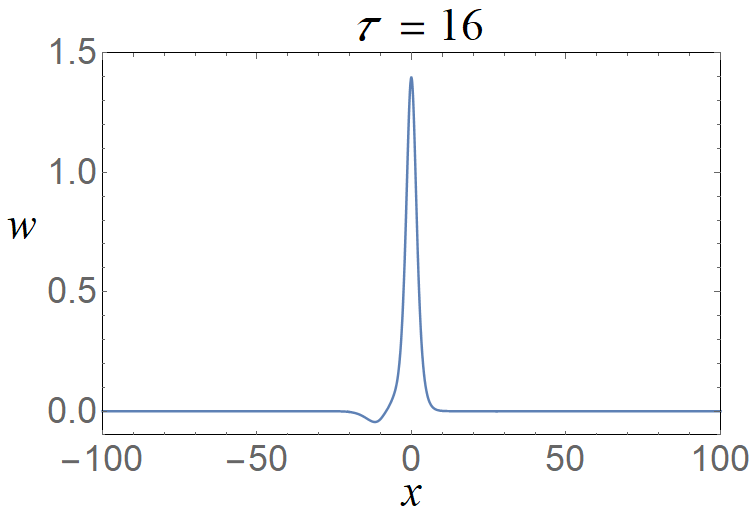}
\hspace{0.3cm}%
\includegraphics[scale=0.25]{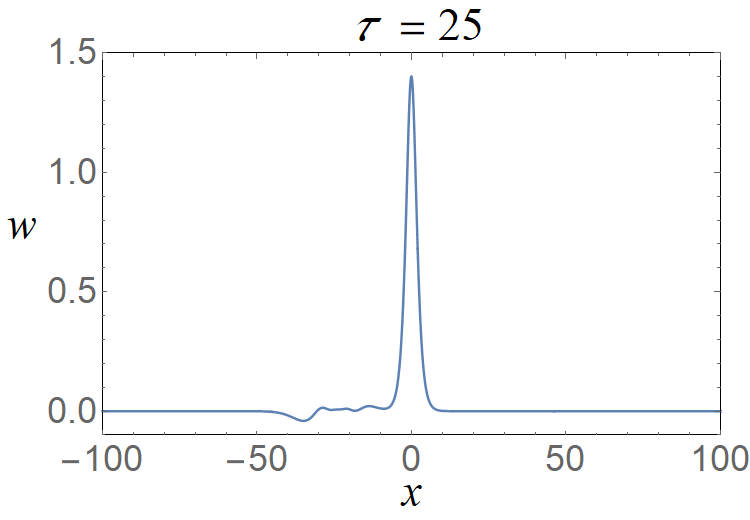}
\caption{The interaction of the solitary wave with $v=1.7$ with the
localized inhomogeneity introduced by Eq. \eqref{F} in the co-moving frame.
The inhomogeneity is initially set at $x=20$. Up to $\protect\tau =10$, the
form of the wave remains practically intact. At $\protect\tau =16$ we
observe deformation of the wave, and at $\protect\tau =25$ we observe the
emergence of a small-amplitude shelf which, along with the accompanying
tail, is attached to the (primary) solitary wave.}
\label{fig:SingleSoliton_forced_v_1_7}
\end{figure}

Naturally, there remains an important open question whether the mass-fitting
factor, used for the analysis of the of solitary-wave collisions, has some
extension/connection to the mass variation in the wave-defect interaction.
However, given the different nature of the two interactions (and the fact
that we can only pursue the solitary-wave interaction until the substantial
emission of radiation occurs, as explained in Figs.~\ref{fig:comparison1}-%
\ref{fig:SingleSoliton_forced_v_1_7}), this issue stays outside the purview
of the present work.

\end{document}